\documentclass[a4paper]{aa}
\usepackage{graphicx}
\usepackage{natbib}
\usepackage{longtable,lscape}
\usepackage{slashbox}

\usepackage{graphicx}
\usepackage{txfonts}
\usepackage{natbib}
\begin{document}
\title{Photocentric variability of quasars caused by variations in their inner structure: Consequences for Gaia measurements}

\subtitle{}

\author{L.\v C. Popovi\'c\inst{1,2} \and P. Jovanovi\'c\inst{1,2} \and
M. Stalevski\inst{1,2,8} \and S. Anton\inst{3,4} \and A. H.
Andrei\inst{5,6,7}  \and J. Kova\v cevi\'c\inst{1,2} \and M.
Baes\inst{8}}

\institute{Group for Astrophysical Spectroscopy, Astronomical Observatory, Volgina 7, 11060 Belgrade 74,
Serbia 
\and
Isaac Newton Institute of Chile, Yugoslavia Branch, Serbia
\and
CICGE, Faculdade de Ci\^encias da Universidade do Porto, Portugal
\and
SIM,  Faculdade de Ci\^encias da Universidade de Lisboa, Portugal
\and
Observat\'orio Nacional/MCT, R. Gal. Jos\'e Cristino 77, CEP 20921-400 Rio de
Janeiro, Brazil
\and
Osservat\'orio Astronomico di Torino/INAF, Strada Osservat\'orio 20, 10025 Pino
Torinese, Italy
\and
SYRTE/Observatoire de Paris, 61 Avenue de l�Observatoire, 75014 Paris, France
Obswervat�rio do Valongo/UFRJ, Ladeira Pedro Ant\'onio 43, CEP 20080-090 Rio de
Janeiro, Brazil
\and
Sterrenkundig Observatorium, Universiteit Gent, Krijgslaan 281-S9,
Gent, 9000, Belgium}

\titlerunning{Photocentric variability of AGNs}

\authorrunning{L.\v C. Popovi\'c et al.}

\offprints{L.\v C. Popovi\'c, \\ \email{lpopovic@aob.rs}\\ }

\date{Received May 12, 2011; accepted  --- ---, 2011}


\abstract
{We study the photocenter position variability caused by variations in
the quasar inner structure. We consider the variability in the accretion
disk emissivity and torus structure variability caused by the different
illumination by the central source. We discuss the possible detection of
these effects by Gaia. Observations of the photocenter variability
in two AGNs, SDSS J121855+020002 and SDSS J162011+1724327 have been 
reported and discussed.}
{For variations in the quasar inner structure,
we explore how much this effect can affect the position
determination and whether it can   (or not) be  detected with the Gaia mission.}
{We use   models of (a) a relativistic disk, including the
perturbation that can increase the brightness of  part of the disk, and
consequently offset the photocenter position, and (b)   a
dusty torus that absorbs and re-emits the incoming radiation from
the accretion disk (central continuum source). We estimate the value
of the photocenter offset caused by these two effects. }
{We found that perturbations in the inner structure can cause a
significant offset to the photocenter. This offset depends on the
characteristics of both the perturbation and accretion disk and on the structure of
the torus. In the case of the two considered QSOs, the observed
photocenter offsets cannot be explained by variations in the
accretion disk and other effects should be considered. We discuss the
possibility of exploding stars very close to the AGN source, and
also  that there are two variable sources at the center
of these two AGNs that may indicate a binary supermassive black
hole system on a kpc (pc) scale.}
{ The Gaia mission seems to be very promising, not only for
astrometry, but also for exploring the inner structure of AGNs. We
conclude that variations in the quasar inner structure can affect
the observed photocenter (by up to several mas). There is a chance to
observe such an effect in the case of bright and low-redshift QSOs. }


\maketitle
%

\section{Introduction}

Gaia is a global astrometric interferometer mission
that aims  to determine high-precision astrometric
parameters for one billion objects with apparent magnitudes in the
range 5.6 $\le V\le 20$ \citep[see e.g.][]{pe01,li08}. It
is foreseen that  500 000 QSOs (quasi-stellar objects) will be among these objects. These QSOs will be used to
construct  a dense optical QSO-based celestial reference frame
\citep[see][]{bo10}. The relevance of QSOs to the celestial frames
compliant to the ICRS, such as the current ICRF2 or the Gaia celestial
reference frame, relies on a photocenter position stability at the
sub-mas level.  Sub-mas accuracy in the measured positions is the goal of Gaia,
namely for objects of 12 mag around  0.003 mas, of 15 mag 0.01 mas and 20
mag 0.2 mas  \citep[][]{pe01}. 

However, QSOs are active galactic
nuclei (AGNs) in whose central region  different physical
processes occur that may cause a variation in the photometric
center of the object. According to the standard model of AGNs, the
central region of a QSO consists of a SMBH
($10^7-10^{10}\ M_\odot$) surrounded by an accretion disk
\citep[see][]{sul00}, and a broad emission-line region (BLR). That
central region might be surrounded by dust, arranged in a
toroidal-like distribution. All these components radiate, and its strength
is a function of the geometry of the system, and  its orientation
relative to the observer. One of
the most important  properties of AGNs is their flux variability, which may have  multiple origins such as  variation in the accretion rate,  instabilities of the accretion disk
around the central black hole, supernova bursts, jet instabilities, and
gravitational microlensing \citep[see e.g.][]{a09,sh10,p11}.

{ \cite{t11} reported on the possibility of a correlation
between the flux variability  and photocenter motion in QSOs, which is  a very
relevant subject for missions such as Gaia. There are different sources
for photocenter variation.  It is well-known that the
main output of the different structures of an AGN (such as accretion disk, jets, line-emitting regions, torus, etc.) differ in energy,
consequently the sizes and position of the emitting regions are
"wavelength dependent''. Opacity effects also explain the frequency-dependent core-shifts in the radio synchrotron emission at the base
of relativistic jets \citep{po09}, and core shifts (between two radio wavelengths) of up to  1.4 mas have
been reported by \cite{kov08}.}

As we mentioned above, {an AGN has a complex structure, and one can expect that the origin of this
variation is caused by the inner structure of this object, as for instance  a torus that is illuminated by a varying central continuum
may contribute to some  photocenter variation. However, variable processes occurring in the accretion disk, such as outburst, and
perturbations \citep[see e.g.][]{jov10,p11}.} 

In this paper, we investigate the spectro-photocentric variability of
quasars caused by changes in their inner structure. We consider: (a)
a perturbation in a relativistic accretion disk around a SMBH, and  (b)
changes in the pattern of radiation scattered by the dust
particles in the surrounding torus caused by the variations in the
accretion disk luminosity and dust sublimation radius.

The aims of the paper are: (a) to show how much these effects may
contribute to the variability of the photocenter, i.e. to quantify
``noise`` and more accurately characterize any resulting error in the
position determination; (b) to estimate the possibility of detecting
this effect with Gaia mission; and (c) to identify in which QSOs these
effects may be dominant.

The paper is organized as follows: in \S\ref{sec:disk} and
\S\ref{sec:tor} we present the models and
parameters of the accretion disk and dusty torus; in \S\ref{sec:res}
results of our simulations are given for different parameters of both the disk
and torus and at different redshifts; in \S\ref{sec:obs}, we consider
the properties of two quasars in the context of obtained results from
our simulations; and in \S\ref{sec:conc}, we outline our conclusions.

In this paper, we use a flat cosmological model with the following
parameters:
$\Omega_\mathrm{m}$­ = 0.27, $\Omega_\mathrm{\Lambda} = 0.73$, and
$H_0 = 71$ km s$^{-1}$ Mpc$^{-1}$.

\section{Disk model around SMBH}
\label{sec:disk}

In the standard model of an AGN accretion disk, accretion occurs via an
optically thick and geometrically thin disk. The effective optical
depth in the disk is very high and photons are close to thermal
equilibrium with electrons \citep{jov09}. The spectrum of
thermal radiation emitted from the accretion disk surface depends on its
structure and temperature, hence on the distance to the
black hole.

An accretion disk around a supermassive black hole at the center of
an AGN extends from the radius of a marginally stable orbit $R_{ms}$ to
 several thousands of gravitational radii. On the basis of 
radiation emitted in different spectral bands, it can be stratified
in several parts \citep{jov09}: a) an innermost part close to the
central black hole that emits X-rays and  extends from the
radius of marginally stable orbit $R_{ms}$ to  several tens of
gravitational radii; b) a  central part ranging from $\sim 100\ R_g$
to $\sim 1000\ R_g$, which emits UV radiation; and c) an outer part
extending from several hundreds to several thousands $R_g$, from
which the optical emission orriginates \citep{Eracleous94,Eracleous03}.

Here we  consider an optical emission disk. The model is
described in our previous papers \citep[see
e.g.][]{lcp03,jov09,jov10}, and here will not be repeated in
detail. We model the emission from an accretion disk using numerical
simulations based on a ray-tracing method in a Kerr metric \citep[see
e.g.][and references therein]{jov09}. In this method, one divides the
image of the disk on the observer's sky into a number of small
elements (pixels), and for each pixel the photon trajectory is
traced backward from the observer by following the geodesics in a
Kerr space-time. Although this method was developed for studying the
X-ray radiation originating from the innermost parts of the
disk close to the central black hole \citep[see e.g.][]{pj08}, it
can be also successfully applied to the modeling of the UV/optical
emission  originating from the outer regions of the disk
\citep[see e.g.][]{jov10}. 

{However, some general
relativistic and strong gravitational effects (such as gravitational
redshift) are significant only in the innermost regions of the
accretion disk, close to its marginally stable orbit. Since the
inner radius of the disk is here taken to be $100\ R_g$, and it
has  a small inclination angle, these effects will have
a negligible influence on the photocenter displacement. Therefore, we
assumed a non-rotating central black hole, such that the
Kerr metric reduces to its special Schwarzschild case. In this way,
we also included in our simulations some Newtonian and special
relativistic phenomena, such as the Doppler effect and relativistic
beaming (see Fig. \ref{fig01} and top left panel of Fig. \ref{fig02}), which cannot be
neglected even at such relatively large distances from the central
black hole,  hence  could cause significant displacements of
the photocenter from the position obtained by simply averaging
the assumed emissivity function of the disk. On the other hand, this
method is very convenient for investigating  photocenter
variability, because the relativistic ray-tracing enables us to
calculate for instance the brightness of each pixel in the
accretion disk image on the observer's sky. This disk image can
 then be used to easily obtain the photocenter position, as we
 show in the later text.}

\subsection{The model of a bright spot--like perturbing region}

\begin{figure}
\centering
\includegraphics[width=\columnwidth]{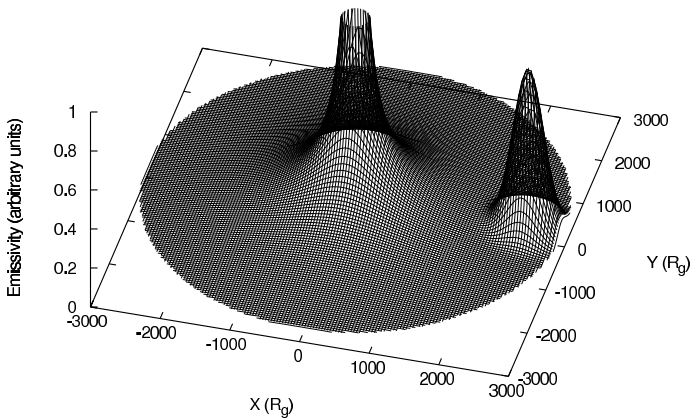}
\caption{A 3D plot of modified disk emissivity given
by Eq.~(\ref{eq1}) for $100\ \mathrm{R_g}\le r(x,y) \le 3000\
\mathrm{R_g}$, $q=-1$ and for the following parameters of perturbing
region: $\varepsilon_p=10$, $x_p=2500\ \mathrm{R_g}$,
$y_p=0\ \mathrm{R_g}$, and $w_p = 300\ \mathrm{R_g}$.}
\label{fig01}
\end{figure}

\begin{figure}
\centering
\includegraphics[width=\columnwidth]{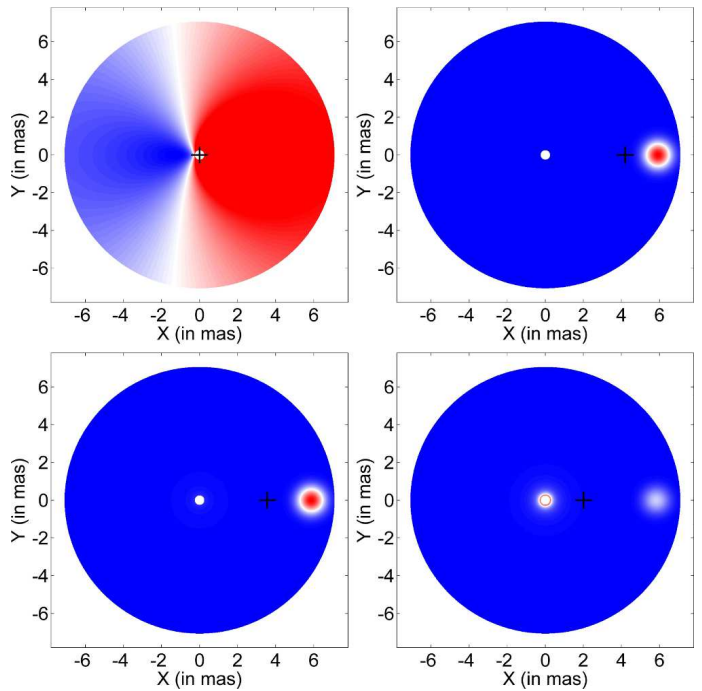}
\caption{{Simulations of the accretion disk without (top left) and
with perturbation for three different values of
emissivity index: $a=0$ (top right), $a=-1$ (bottom left), and $a=-2$ (bottom right).
The photocenter positions are denoted by crosses. In the top left panel, color represents the energy
shift due to relativistic effects (i.e. ratio of the observed to emitted energy),
while in the other three panels it represents the observed flux (in arbitrary units).}
The inner and outer radii of the disk, as well as the position and
width of the perturbing region, are the same as in Fig.~\ref{fig01}.
The maximum emissivity of the perturbing region is taken to be ten
times greater than the emissivity of the disk at its inner radius.
Linear distances are converted to angular units along the $x$ and
$y$ axes assuming an accretion disk located at cosmological redshift
$z=0.01$ around the central black hole with mass of 10$^{10}\ M_\odot$.}
\label{fig02}
\end{figure}

To model a bright spot on the disk, we considered
perturbations in the surface emissivity on some region of the disk.
Surface emissivity of the disk is usually assumed to vary with
radius as a power law \citep[e.g.][]{lcp03} $\varepsilon (x,y) =
\varepsilon _0  \cdot r^a(x,y),$ where $\varepsilon _0$ is an
emissivity constant, $a$ is emissivity index, and $(x,y)$ is a
position along the disk. We introduce the perturbation in accretion
disk emissivity (bright spot) in the form of a two-dimensional circular
Gaussian, by modifying its power-law emissivity according to \citep{jov09}
\begin{equation}
\varepsilon' (x,y) = \varepsilon (x,y)
\cdot \left( {1 + \varepsilon _p
\cdot e^{ - \left( {\left( {\frac{{x -x_p}}{{w_p }}} \right)^2  +
\left( {\frac{{y - y_p }}{{w_p }}} \right)^2 }
\right)} } \right),
\label{eq1}
\end{equation}
where $\varepsilon' (x,y)$ is the modified disk emissivity,
$\varepsilon (x,y)$ is the ordinary power-law disk emissivity at
the same position $(x,y)$, $\varepsilon_p$ is the emissivity of the
perturbing region (i.e. the amplitude of the bright spot), $(x_p,y_p)$ is
the position of the perturbing region with respect to the disk center
(expressed in gravitational radii, hereafter denoted by $R_g=GM/c$,
where $M$ is the mass of the SMBH, and
G and c are well known constants) and
$w_p$ is its width (also in $R_g$). A three-dimensional (3D) plot of
the above expression for the modified emissivity law is given in Fig.
\ref{fig01}.

Owing to relativistic effects, photons emitted from the disk at frequency
$\nu_{em}$ will reach observers at infinity at frequency $\nu_{obs}$,
and their ratio determines the shift caused by these effects $g =
\nu _{obs}/\nu _{em}$. The total observed flux at the observed
energy $E_{obs}$ is then given by
\begin{equation}
F\left( {E_{obs}}  \right) = {\int\limits_{image}
{\varepsilon'(x,y)}}\, g^{4}\delta \left( {E_{obs} -
gE_{0}}  \right)d\Xi ,
\label{eq2}
\end{equation}
where $\varepsilon' \left( {r} \right)$ is the modified disk
emissivity given by Eq.~(\ref{eq1}),
$d\Xi$ is the solid angle subtended by the disk in the observer's
sky, and $E_{0}$ is the rest energy.

This simple model is suitable for our purpose because it
allows us to change the amplitude, width, and location of bright spots with
respect to the disk center. In this way, we are able to simulate
the displacement of a bright spot along the disk, and its widening and
amplitude variations with time.

\begin{table}
\centering
\caption{The simulated offsets of photocenter (in mas)
caused by perturbation to the accretion disk emissivity for different
values of its redshift and mass of a central black hole. Other
parameters correspond to the bottom left panel of Fig.~\ref{fig02}.}
\begin{tabular}{|c|c|c|c|c|c|}
\hline
$M_{BH}$& \multicolumn{5}{c|}{$z$} \\
\cline{2-6}
$(M_\odot)$& 0.01 & 0.05 & 0.10 & 0.15 & 0.20 \\
\hline
\hline
10$^8$ & 0.036 & 0.007 & 0.004 & 0.003 & 0.002 \\
\hline
10$^9$ & 0.355 & 0.074 & 0.039 & 0.028 & 0.022 \\
\hline
10$^{10}$ & 3.550 & 0.744 & 0.394 & 0.278 & 0.220 \\
\hline
\end{tabular}
\label{tab01}
\end{table}

\subsection{Modeled offset of the photocenter caused by a perturbation (bright spot)
in the disk}

The observed photocenter $(X_{pc}, Y_{pc})$ of
the accretion disk can be
modeled as a centroid of observed flux $F(E_{obs})$ over the disk
image, i.e. as the mean of impact parameters $(x, y)$ of all pixels
along the disk image, weighted by $F$
\begin{eqnarray}\label{eqn:phc}
X_{pc}=\frac{\sum_{i=1}^{N} \sum_{j=1}^{N}
F(i,j)\cdot x(i,j)}{\sum_{i=1}^{N} \sum_{j=1}^{N} F(i,j)},
\nonumber \\
Y_{pc}=\frac{\sum_{i=1}^{N} \sum_{j=1}^{N}
F(i,j)\cdot y(i,j)}{\sum_{i=1}^{N} \sum_{j=1}^{N} F(i,j)},
\end{eqnarray}
where $(i,j)$ is a point on a $N\times N$ grid of the
disk image pixels.

We consider a perturbation (or bright spot) at a certain part of the
disk, for different values of the spot brightness, and calculate the
photocenter. In Fig. \ref{fig02}, we present
simulations of the photocenter variability.

\subsection{Parameters of the disk and perturbation (bright spot)}

In the model, we are able to change the parameters of the accretion
disk (dimension, emissivity, inner and outer radius, inclination)
and the parameters of the perturbation (size, position, and
brightness). Taking into account the results of previous studies, one can
expect  the dimensions of the accretion disk to be several thousands of
gravitational radii \citep[see e.g.][]{Eracleous94,Eracleous03,p11},
hence here we assume an accretion disk with an inner and outer radius
of R$_{inn}$ = 100 R$_g$ and R$_{out}$=3000 R$_g$, respectively.
{In our simulation, we consider a low-inclined ($i=5^\circ$) or near
face-on disk, because of from   investigations of the broad line shapes
a near face-on disk is preferred \citep[see e.g.][]{p04,b09}.
Although the adopted inclination angle is small, it is sufficient to induce
Doppler and relativistic beaming effects
\citep[see e.g. Fig. 9 in][and the corresponding discussion below]{reyn03}.
As shown in \citet{reyn03}, even in the case of a nearly face-on disk,
these effects can still produce  rather broad emission lines, unlike the
case
of a face-on Newtonian disk, which would display  very narrow lines. In addition,
for a steep disk emissivity where $a < -2$, the line emission of the disk is
dominated by its
inner regions $R_{out} < 50\ R_g$. However, for the disk emissivity where $a
>-2$,
the bulk of the line emission comes from the outer regions of the disk, thus both Doppler and relativistic
beaming effects cannot be neglected even at such relatively large distances
from the central
black hole. Since the most realistic values for the emissivity of the disk
are
probably between 0 and $-2$
\citep[see e.g.][]{Eracleous94,Eracleous03,p04,bb09}, we
modeled the disk emissivity index as $a=0$, $a=-1$, and $a=-2$.}

In our simulation, the dimensions of the perturbation (bright
spot) is around 100 -- 300 gravitational radii \citep[see][]{jov10},
taking different values for the brightness and position along the
disk.

\section{Dusty torus model}
\label{sec:tor}

According to the AGN unification model, the central continuum source
is surrounded by the geometrically and optically thick toroidal
structure of dust and gas with an equatorial visual optical depth
much larger than unity. To prevent the dust grains from
being destroyed by the hot surrounding gas, it has been suggested
\citep{krolikbegel88} that the dust in the torus is organized into a large
number of optically thick clumps. In an edge-on view, this dusty torus
blocks the radiation coming from the accretion disk and BLR and
object appears as type 2 active galaxy. When the line of
sight does not cross the dusty torus, both the accretion disk and BLR
are exposed and the object is classified as a type 1 active galaxy.
This dusty torus absorbs the incoming radiation and re-emits it,
mostly in the infrared domain, but  a part of the radiation is also
scattered in the optical domain.

The model of a torus that we used in this work is described in detail in
\cite{stal11}; here we  present only its most
important properties. Our approach allows us to model the torus as a
3D structure,  composed of (a) isolated clumps, or (b)  a two-phase
medium with high-density clumps and low density medium filling the
space between the clumps. We employed a 3D Monte Carlo radiative
transfer code called SKIRT \citep[for more details
see][]{baes03,baes11} to calculate spectral
energy distributions (SED) and images of the torus at different
wavelengths.
\begin{figure*}
\centering
\includegraphics[height=0.42\textwidth]{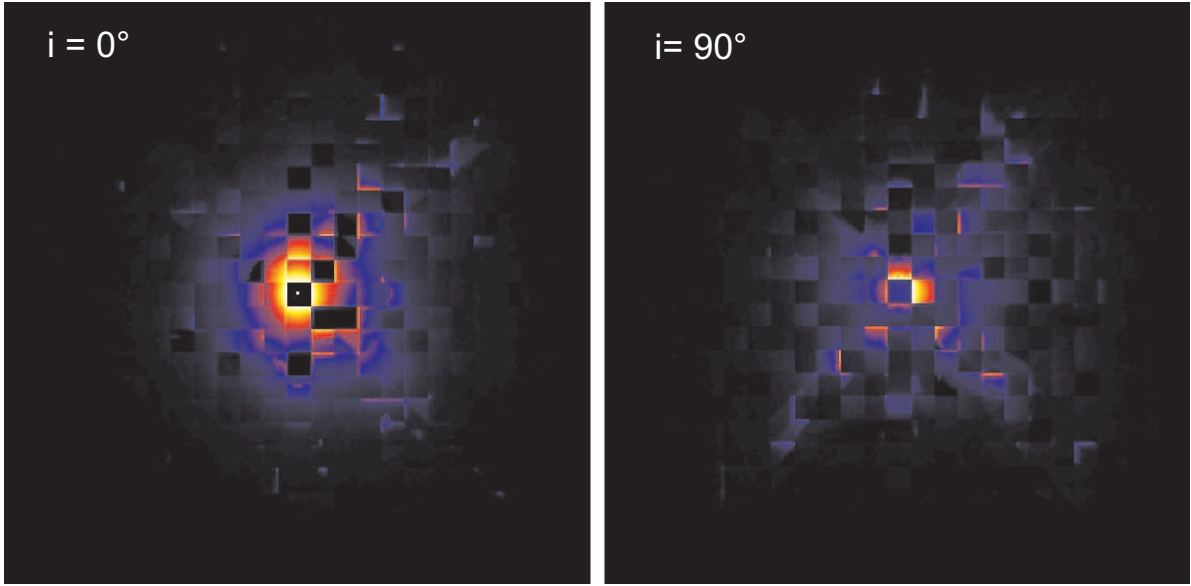}
\caption{Images of torus for face-on (left panel) and edge-on view
(right panel), at $9.7$ $\mu$m, in logarithmic scale. The values of
torus parameters are: optical depth $\tau_{9.7}=5$, dust distribution
parameters $p=1$ and $q=0$, half opening angle $\theta=50^\circ$,
inner radius $R_{in}=0.5$ pc, outer radius $R_{out}=15$ pc; filling
factor $0.25$, contrast $10^9$, size of clumps $1.2$ pc. Luminosity
of the central continuum source is $L=10^{11}L_{\odot}$.}
\label{fig:i0i90}
\end{figure*}
\begin{figure*}
\centering
\includegraphics[height=0.34\textwidth]{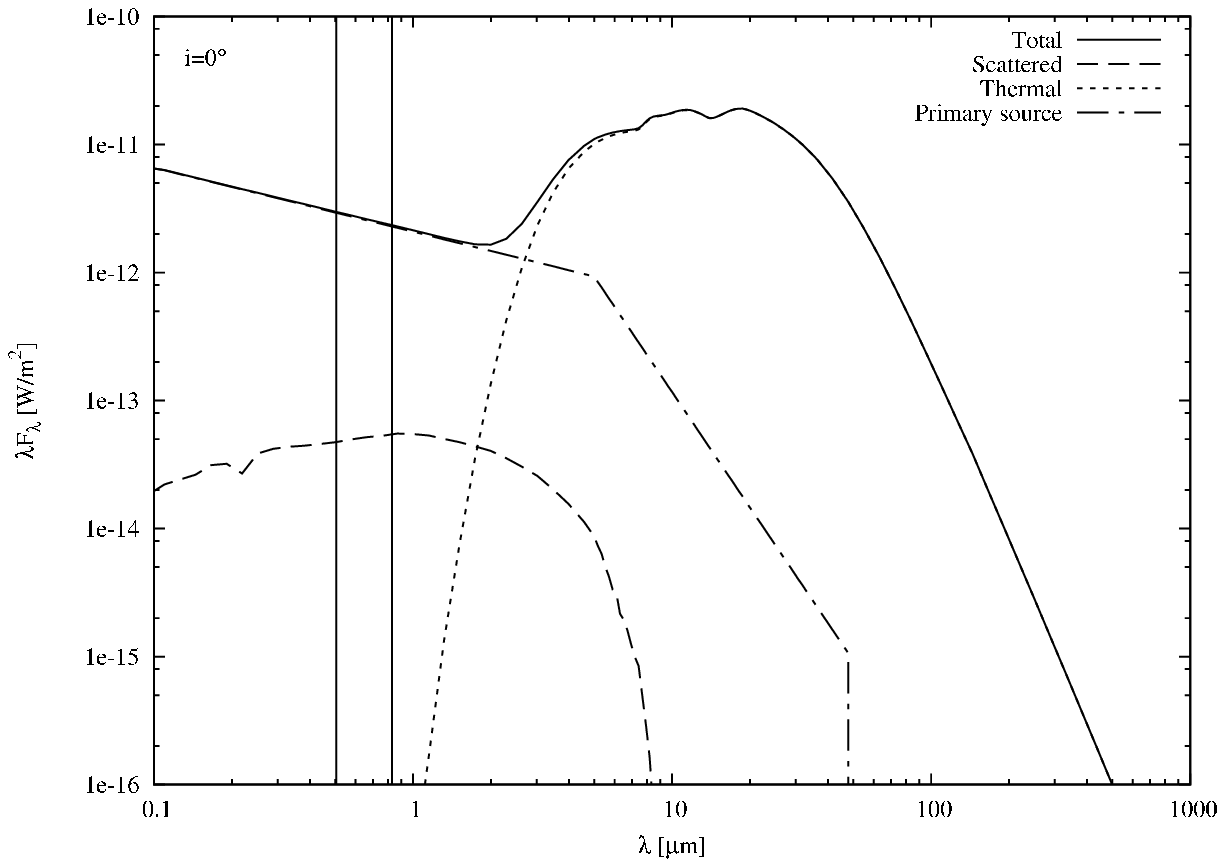}
\includegraphics[height=0.34\textwidth]{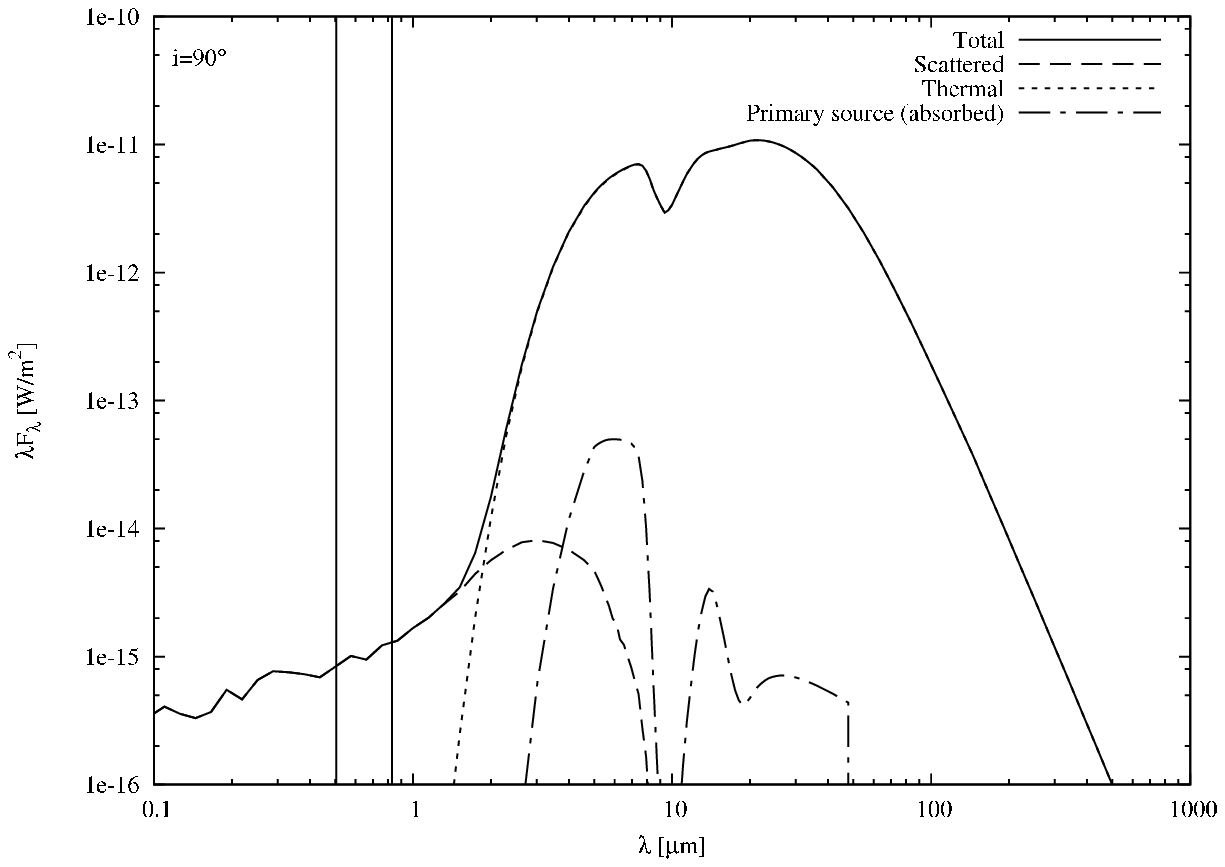}
\caption{The total (solid line), thermal (dotted line), scattered
(dashed line), and primary source (dash-dotted line) emission are
plotted. The left panel is a type 1 inclination ($i=0^\circ$), the
right panel a type 2 inclination ($i=90^\circ$). The two vertical
lines  indicate the central wavelengths of the two dispersing prisms
of the Gaia photometric instrument (integrated with the astrometric
instrument), at $0.50$ and $0.82$ $\mu$m. The values of torus parameters
are the same as taken in Fig. \ref{fig:i0i90}.}
\label{fig:flux}
\end{figure*}

\begin{figure*}
\centering
\includegraphics[height=0.24\textwidth]{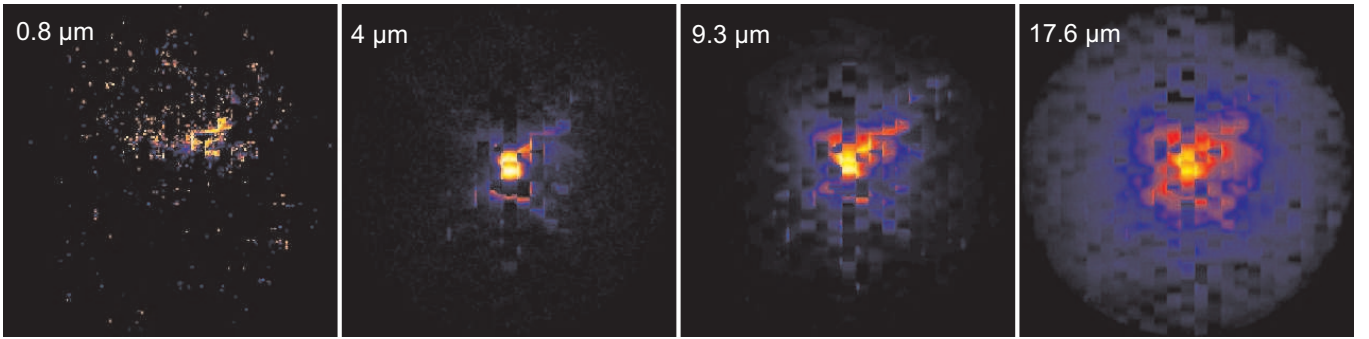}
\caption{Images of the torus at different wavelengths. From left to
right, panels represent model images at $0.83$, $3.98$, $9.31$, and $17.6$
$\mu$m. Images are in logarithmic scale. The visible squared
structure is due to the clumps which in our model are in the form of
cubes. The inclination is $i=50^\circ$; the values of other
parameters are the same as taken in Fig. \ref{fig:i0i90}.}
\label{fig:torimg}
\end{figure*}

We approximate the obscuring toroidal dusty structure with a conical
torus (i.e. a flared disk). Its characteristics are defined by (a)
half opening angle $\theta$, (b) inner and outer radius, $R_{in}$
and $R_{out}$ respectively, and (c) parameters describing the dust
density distribution, $p$ and $q$. The inner radius is calculated
according to the prescription given by \cite{barvainis87}
\begin{equation}\label{eqn:rmin}
R_{in}\simeq 1.3 \cdot \sqrt{L_{46}^{AGN}}\cdot T_{1500}^{-2.8}
\qquad [pc] ,
\end{equation}
where $L_{46}^{AGN}$ is the bolometric ultraviolet/optical luminosity
emitted by the central source, expressed in units
of $10^{46}$ erg s$^{-1}$ and $T_{1500}$ is the sublimation
temperature of the dust grains given in units of $1500$ K.

We describe the spatial distribution of the dust density with a law
that allows a density gradient along the radial direction and with
polar angle, similar to the one adopted by \cite{granatodanese94}:
\begin{equation}\label{eqn:dens}
\rho\left(r,\theta \right)\propto r^{-p}e^{-q|cos(\theta)|} ,
\end{equation}
where $r$ and $\theta$ are coordinates in the adopted coordinate
system. The dust mixture consists of separate populations of graphite
and silicate dust grains with a classical MRN size distribution
\citep{mrn77}. The total amount of dust is
fixed based on the equatorial optical depth at $9.7$ $\mu$m
($\tau_{9.7}$).

To generate a clumpy medium, we apply the algorithm described by
\cite{wittgordon96}. The parameters that define the clumpiness of the dusty
medium are the filling factor and contrast. Filling factor sets the number
of clumps; contrast is defined as the ratio of the dust density in
the high- to low-density phase. For example, setting the contrast to
unity would result in a continuous, smooth dust distribution. Setting
an extremely high value of contrast ($>$1000) effectively puts all the
dust into the clumps, without a low-density medium between them.

\subsubsection{Spectral energy distribution of the primary continuum
source}

The primary continuum source of dust heating is the intense
UV-optical continuum coming from the accretion disk. A very good
approximation of its emission is a central, point-like energy source,
emitting isotropically. Its SED is very well-approximated by a
composition of power laws with different spectral indices in
different spectral ranges. The adopted values are:
\begin{equation}\label{eqn:source}
\lambda L(\lambda)\propto\left\{
\begin{array}{lrrr}
\lambda^{1.2}  &  \; 0.001 < \lambda < 0.01  & [\mu m]\\
\lambda^{0}    &  \; 0.01  < \lambda < 0.1   & [\mu m]\\
\lambda^{-0.5} &  \; 0.1   < \lambda < 5     & [\mu m]\\
\lambda^{-3}   &  \; 5     < \lambda < 50    & [\mu m].
\end{array}
\right.
\end{equation}
These values have been quite commonly adopted in the literature, and
come from both observational and theoretical arguments
\citep[see e.g.,][]{schartmann05}.

\subsection{Modeled offset of the photocenter due to  variations
in the central source luminosity and dust sublimation radius}
\label{sec:pcofftor}

Variations in the primary continuum source-emission (not only
perturbations, but  also changes to the total luminosity of accretion
disk) may also cause a photocenter offset due to another effect.
According to  Eq.~(\ref{eqn:rmin}), the dust sublimation radius (i.e.
inner radius of torus)  depends on the total bolometric
luminosity of the central source (accretion disk). Thus, with
increasing central source luminosity, the inner radius of the torus also
increases. This means that (a) the innermost structure of the
torus changes and (b) the radiation from the central source is able to
penetrate further into the torus. These two effects will change the
illumination of  clumps and the pattern of the scattered radiation, which may
lead to  variations in the photocenter position. The photocenter of
dusty torus is calculated in the same way as for the accretion disk (see
Eq.~\ref{eqn:phc}).

\subsection{Parameters of the dusty torus model}
\label{sec:torpar}

The parameter that has a very prominent effect on the shape of SED
is the inclination. The inclination $i=0^\circ$ corresponds to
a face-on (type 1) AGN and $i=90^\circ$ an edge-on (type 2) AGN.
{Fig. \ref{fig:i0i90} shows images of a torus model for a face-on and
an edge-on view.} In Fig. \ref{fig:flux}, we present the total SED and
its thermal and scattered components, along with primary source SED,
for these two inclinations. As it can be seen from this figure, there
is a clear distinction between the cases of a dust-free line of sight
($i=0^\circ$; left panel) and those that pass through the torus
($i=90^\circ$, right panel). In the case of dust-free lines of
sight, we can directly see the radiation coming from the accretion disk,
while in the case of dust-intercepting paths most of the radiation
is absorbed and re-emitted at different wavelengths. From the
figure, one can also see that the thermal component predominates the mid-
and far-infrared parts of a SED and its shape is similar for both
face-on and edge-on orientations. However, the shape and amount of the
scattered component is quite different;  in the edge-on view, it
determines the total SED shortward of $1$ $\mu$m, while in the face-on
view it is negligible compared to the primary source emission. We
illustrate this further in Fig.~\ref{fig:torimg}, where images of the
torus at different wavelengths are presented. Shortward of $1$
$\mu$m (first panel), the thermal component is negligible and only the
scattered component that arises randomly from the entire torus is
present. In the near- and mid-infrared domain (second and third
panel), the thermal radiation from the inner (and hotter)
region predominates. At longer wavelengths (forth panel), emission
arises from the dust placed further away.

Since in the wavelength range relevant to this work ($<1$ $\mu$m), the
scattered component of dust emission is dominant, the other
parameters, (e.g. those defining geometry and dust distribution)
have only a marginal influence on images of the torus. Therefore, we fix
the following values of torus parameters: optical depth
$\tau_{9.7}=5$; dust distribution parameters (see Eq~\ref{eqn:dens})
$p=1$ and $q=0$; half opening angle $\theta=50^\circ$; and outer radius
$R_{out}=15$ pc. For the parameters defining clumpiness, we adopt a filling factor of $0.25$, which allows single
clumps as well as clusters of several merged clumps, and to define  the
contrast an extremely high value ($10^9$), which effectively
puts all the dust into the clumps, without any dust being smoothly
distributed between the clumps. For the size of clumps, we adopt the
value of $1.2$ pc. We calculated models at two inclinations,
$i=30^\circ$ (dust-free line of sight) and $i=50^\circ$ (line of
sight that passes through the torus). For the total bolometric
luminosity of the central continuum source, we adopt the values of
$L=1, 3, 6, 10~ \times10^{11}$ $L_{\odot}$. According to 
Eq.~\ref{eqn:rmin} (assuming the dust sublimation temperature of
$1200$ K), the corresponding values of the inner radius of torus
are $R_{in}=0.5, 0.82, 1.16, 1.5$ pc, respectively.

\section{Modeled photocenter offset caused by changes to the inner quasar structure: Results and discussion}
\label{sec:res}

\subsection{Photocenter offset caused by a perturbation
(bright spot) in the accretion disk}

We performed simulations for different emissivities and different
positions of the bright spot on the disk. As an example, we present in Fig.
\ref{fig02} three simulations of the photocenter offset due to a
perturbation in the disk for three different values of its
emissivity index. In Fig. \ref{fig02}, we show the
simulations of accretion disk without (top left panel) and with
a perturbation (other three panels), i.e. the disk images (for a
quasar with a SMBH of 10$^{10}\ M_\odot$ at $z=0.01$) for three
different values of emissivity index $a=0$ (top right), $a=-1$
(bottom left), and $a=-2$ (bottom right). The photocenter positions
are denoted by crosses. The inner and outer radii of the disk are
taken to be 100 and 3000 R$_g$, respectively. The emissivity of the
bright spot is $\varepsilon_p=10$\footnote{ Note here that in the case of tidal disruptions of stars by a supermassive black hole 
the amplification in the total optical brightness  can increase around two times \citep[see][and discussion below]{kom08}, therefore the small bright spot should have a significantly (around one order) higher emissivity than the disk}, the position is $X_p=2500\
\mathrm{R_g}$, $Y_p=0\ \mathrm{R_g}$, and the dimension of the bright
spot is taken to be $w_p = 300\ \mathrm{R_g}$. As  can be seen
from the Figure, the offset of the photocenter depends on the disk
emissivity and it is the most prominent in a disk with flat emission
($q=0$): the corresponding offsets are smaller for steeper
emissivity laws and vice versa. We also note here that we  take  a very
strong perturbation at the disk edge, and that the maximum emissivity of the
perturbing region is taken to be ten times greater than emissivity of
the disk {disk center} (hereafter we refer to this as the central
source).

{Occurrences of perturbations in the accretion disk emissivity
could be caused by several physical mechanisms, such as disk
self-gravity, baroclinic vorticity, disk-star collisions, tidal
disruptions of stars by a central black hole, and fragmented spiral
arms of the disk \citep[see e.g.][and references therein]{jov10}.
All these phenomena appear and last at different frequencies and
timescales, and could cause perturbations of different strengths,
proportions and characteristics. In particular, perturbations of
accretion disk emissivity in the form of flares with high amplitudes are
of great significance because they could provide information about
accretion physics under extreme conditions. The flares with the
highest amplitudes are usually interpreted in terms of tidal
disruptions of stars by supermassive black holes \citep[see
e.g.][and references therein]{kom08}. Stars approaching a SMBH will
be tidally disrupted once the tidal forces of the SMBH exceed the
star's self-gravity, and part of the stellar debris will be
accreted, producing a luminous flare of radiation that persists  on a
timescale of between months and years. This flare is expected to occur in the outer 
part of the disk (similar to our simulations).

 Although, the frequency of these events in
a typical elliptical galaxy is very low, between $10^{-5}$ and
$10^{-4}$ per year \citep[see e.g.][and references therein]{jov10},
 \citet{kom08} reported the discovery of an X-ray outburst
of large amplitude in the galaxy SDSS J095209.56+214313.3, which was
probably caused by the tidal disruption of a star by a supermassive
black hole. Although this was a high-energy (EUV, X-ray) outburst,
its low-energy (NUV, optical, NIR) echo was also detected.}

In general, we found that in the case of luminous bright spot
(smaller than emission in the central source) the offset of the
photocenter will be negligible, especially if the bright spot
appears close to the center. In addition, when there is  high
emissivity in the bright spot close to the central source,
the effect is small. Only a luminous bright spot located relatively
far from the central source can be a good candidate to be observed
with Gaia. To  estimate whether the offset of the photocenter
 can be observed we give numerical values of the photocenter
offsets (in mas) for different redshifts and black hole masses  in
Table~\ref{tab01}. The parameters for the disk and perturbation are
taken as given above, for the emissivity index of $a=-1$.

As  can be seen from Table~\ref{tab01},  the
largest photocenter offsets ($\sim$ several mas) found at  the
lowest redshifts ($z\sim 0.01$) and the most massive
black holes ($M_{BH}\sim 10^{10}\ M_\odot$), where we can expect to find the
accretion disk with the larger dimensions.

\subsection{Photocenter offset due to the variations in both the central
source luminosity and dust sublimation radius}

As explained in \S\ref{sec:pcofftor}, an increase in the accretion disk
luminosity may cause variations in the photocenter position.
Therefore, for the adopted values of torus parameters (see
\S\ref{sec:torpar}) we generated a set of models for different
luminosities and corresponding inner radii (i.e. dust sublimation
radii), i.e. $L=1, 3, 6, 10~
\times10^{11}$ $L_{\odot}$ and $R_{in}=0.5, 0.82, 1.16, 1.5$ pc,
respectively. We calculated models at two inclinations,
$i=30^\circ$ (dust-free line of sight) and $i=50^\circ$ (line of
sight that passes through the torus). For each model, we calculated
the photocenter position and its offset from the one in the starting
model ($L=10^{11}L_{\odot}$).

\begin{figure*}
\centering
\includegraphics[height=0.49\textwidth]{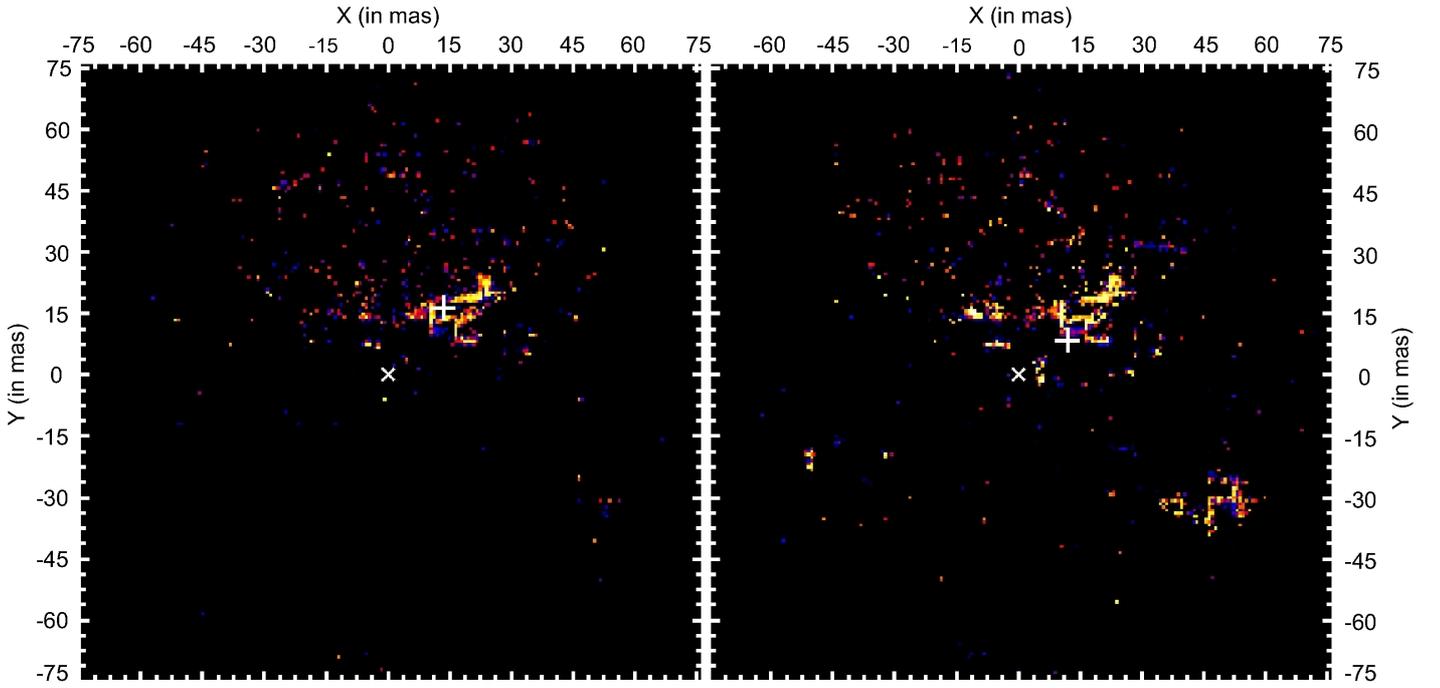}
\caption{Images of torus model at $0.51$ $\mu$m for two different
luminosities and corresponding inner radii, $10^{11}L_\odot$ and
$R_{in}=0.5$ pc (left panel) and $6\times10^{11}L_\odot$ and
$R_{in}=1.16$ pc (right) panel. Photocenter in both panels is
denoted with a white cross; black hole in both panels is at the
center of the images, denoted with 'x' The photocenter offset
between the images is $8.4$ mas. The values of other torus
parameters are the same as in Fig. \ref{fig:torimg}. Images are on a
logarithmic scale.}
\label{fig:torphc}
\end{figure*}

We found that  when the central source is unobscured
($i=30^\circ$), the brightness of the source is dominant and the
photocenter offset is negligible. In Table \ref{tab:tor}, we present
values of the photocenter offset in the case of $i=50^\circ$ and for
different accretion disk luminosities and cosmological redshifts. As
 can be seen from the Table, the photocenter offset is larger for
lower cosmological  redshifts and bigger luminosity outbursts. In
Fig. \ref{fig:torphc}, we present images of the torus in the case of the
largest photocenter offset ($8.4$ mas), at  $z=0.01$, for the
central source luminosities of $10^{11}$ $L_{\odot}$ (left panel)
and $6\times10^{11}$ $L_{\odot}$ (right panel). 

{As  can be seen from Table \ref{tab:tor}, a large jump
in the photocenter offset  between the luminosities of $3$ and
$6$ $~\times10^{11}$ $L_{\odot}$ is present. This is caused by the
change in the illumination of the torus. As the luminosity of the
central source  increases, the inner radius of the torus 
increases as well (the inner structure changes), and the group
of clumps farther away from the center may be illuminated (see  Fig.
\ref{fig:torphc}, right panel). However, a further increase in the
central source luminosity does not change the illumination pattern of
the clumps significantly (depending on the actual distribution of the
clumps) hence the value of the photocenter offset remains nearly
the same. In addition as the central source luminosity continues to  increas, the brightness of the central source begins to dominate, and the
photocenter  gets closer to the central source.}

\begin{table}
\centering
\caption{The simulated offsets of photocenter (in mas)
for different values of redshift and accretion disk luminosity,
calculated for the two photometric instruments with central
wavelengths at $0.50$ $\mu$m and $0.82$ $\mu$m. The values of torus
parameters are the same as in Fig.~\ref{fig:torimg}.}
\begin{tabular}{|c|c|c|c|}
\hline
$L$& \multicolumn{3}{c|}{$z$} \\
\cline{2-4}
$(10^{11}L_\odot)$& 0.01 & 0.05 & 0.10 \\
\hline
\hline
\multicolumn{4}{|c|}{$0.50$ $\mu$m} \\
\hline
3 & 1.579 & 0.208 & 0.039 \\
\hline
6 & 8.400 & 1.886 & 0.860  \\
\hline
10 & 8.170 & 1.353 & 0.693 \\
\hline
\hline
\multicolumn{4}{|c|}{$0.82$ $\mu$m} \\
\hline
3 & 0.814 & 0.252 & 0.135 \\
\hline
6 & 7.120 & 1.422 & 0.990  \\
\hline
10 & 7.978 & 1.466 & 0.843 \\
\hline
\end{tabular}
\label{tab:tor}
\end{table}

\subsection{Photocenter position vs. flux variation}

For one object \cite{t11}  found that a relationship between the
astrometric and photometric variability exists. We also modeled the
expected flux variation with brightness of the perturbed region, and
found that the offset of the photocenter in principle can be a
function of the flux variation only in special cases where there is a
perturbation located at the same place and the brightness 
changes with time. In general, there are many possible
 locations of the perturbations and possible values of their emissivities with
respect to the central source. { The photocenter position varies in terms of both the central source brightness (that may show variability) and  the emissivity of the bright spot, hence the relationship between the astrometric and photometric variability
cannot be assumed as the general rule, although it may exist
particularly in the $\mu$as astrometric regime.}

On the other hand, in the case of the changes in the torus
structure, as can be seen from Table 2, there is a partial
correlation between the photometric and astrometric variability, but
it is not a rule, especially when illumination stays higher.

\begin{figure*}
\centering
\includegraphics[height=0.49\textwidth,angle=-90]{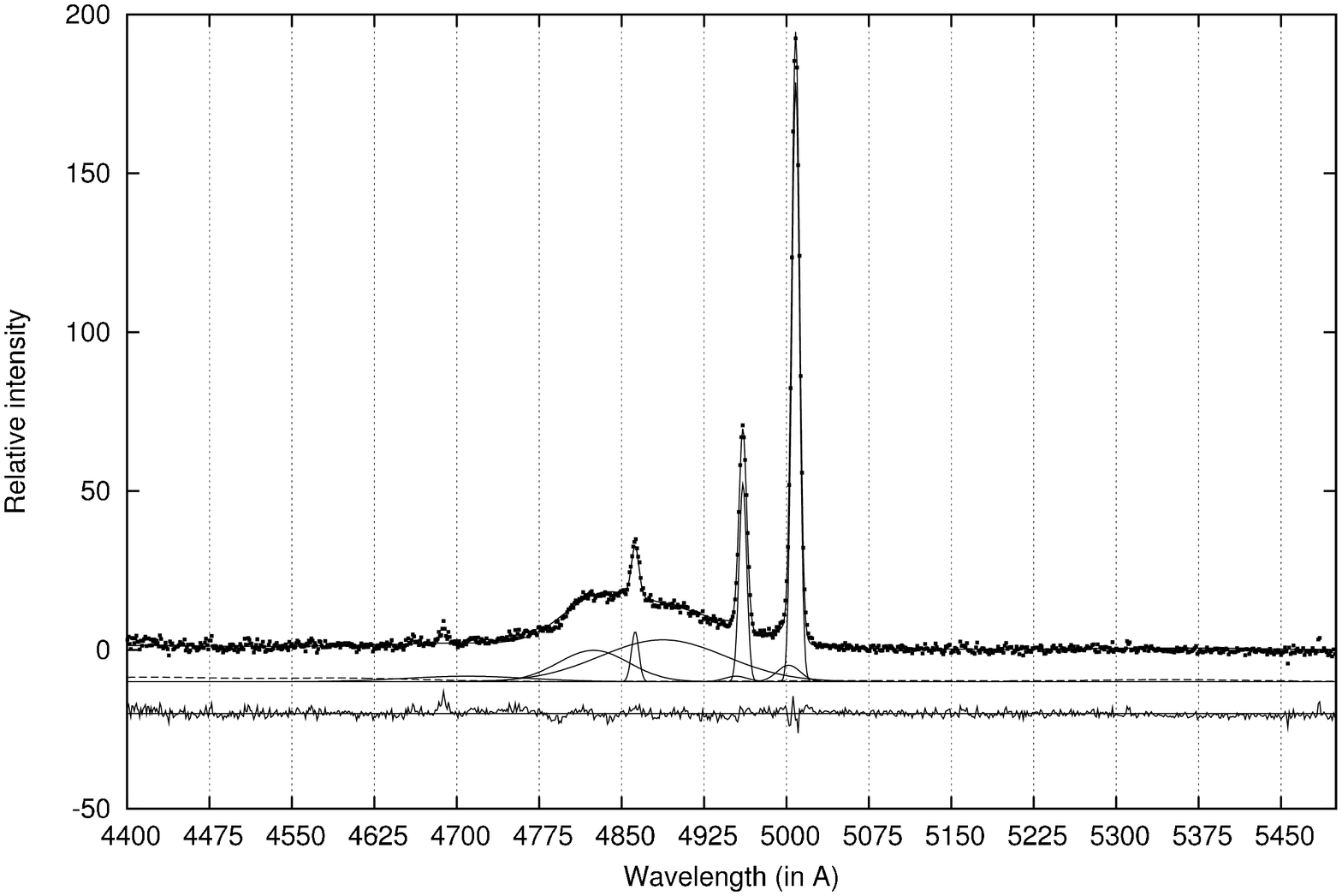}
\includegraphics[height=0.49\textwidth,angle=-90]{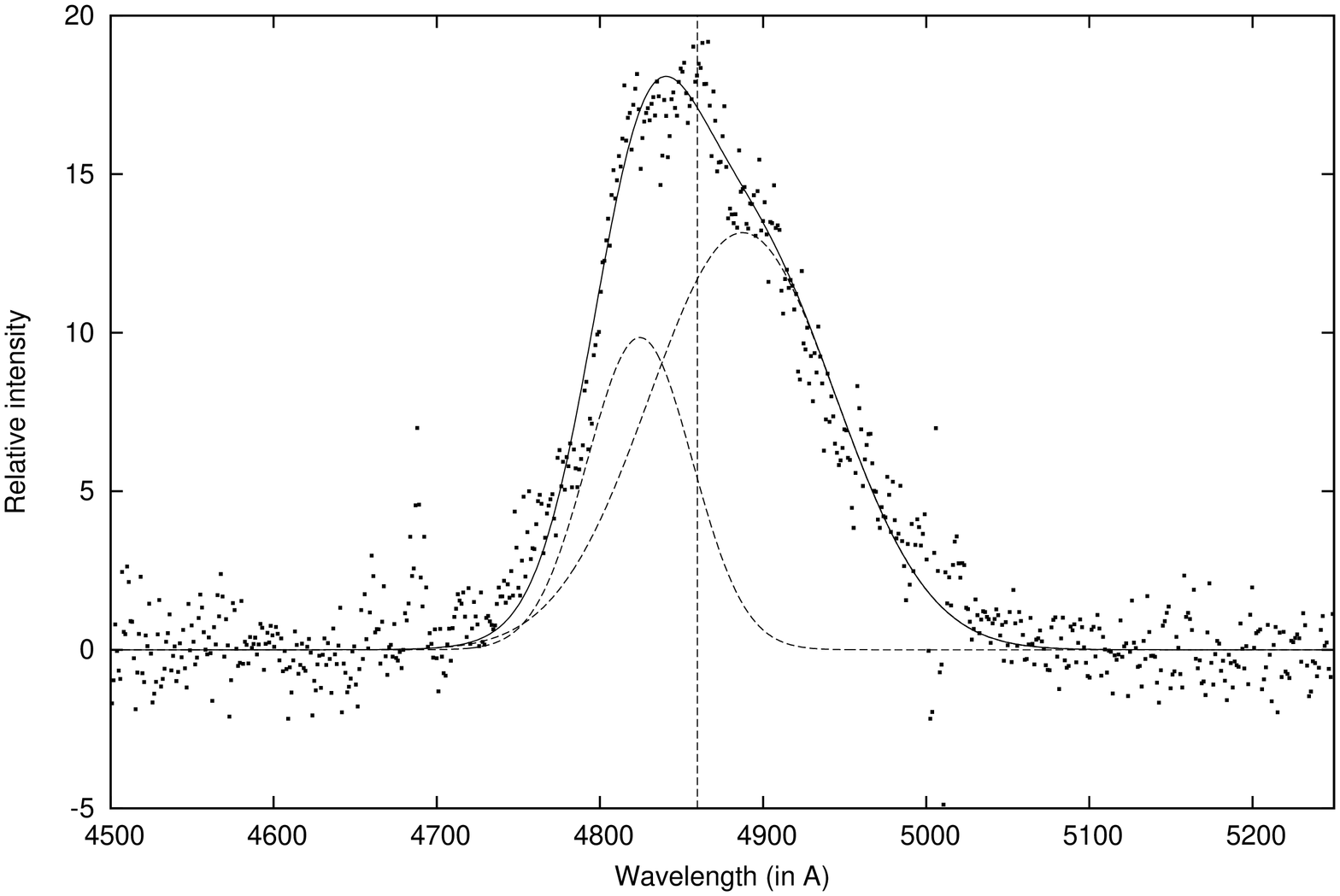} \\
\includegraphics[height=0.49\textwidth,angle=-90]{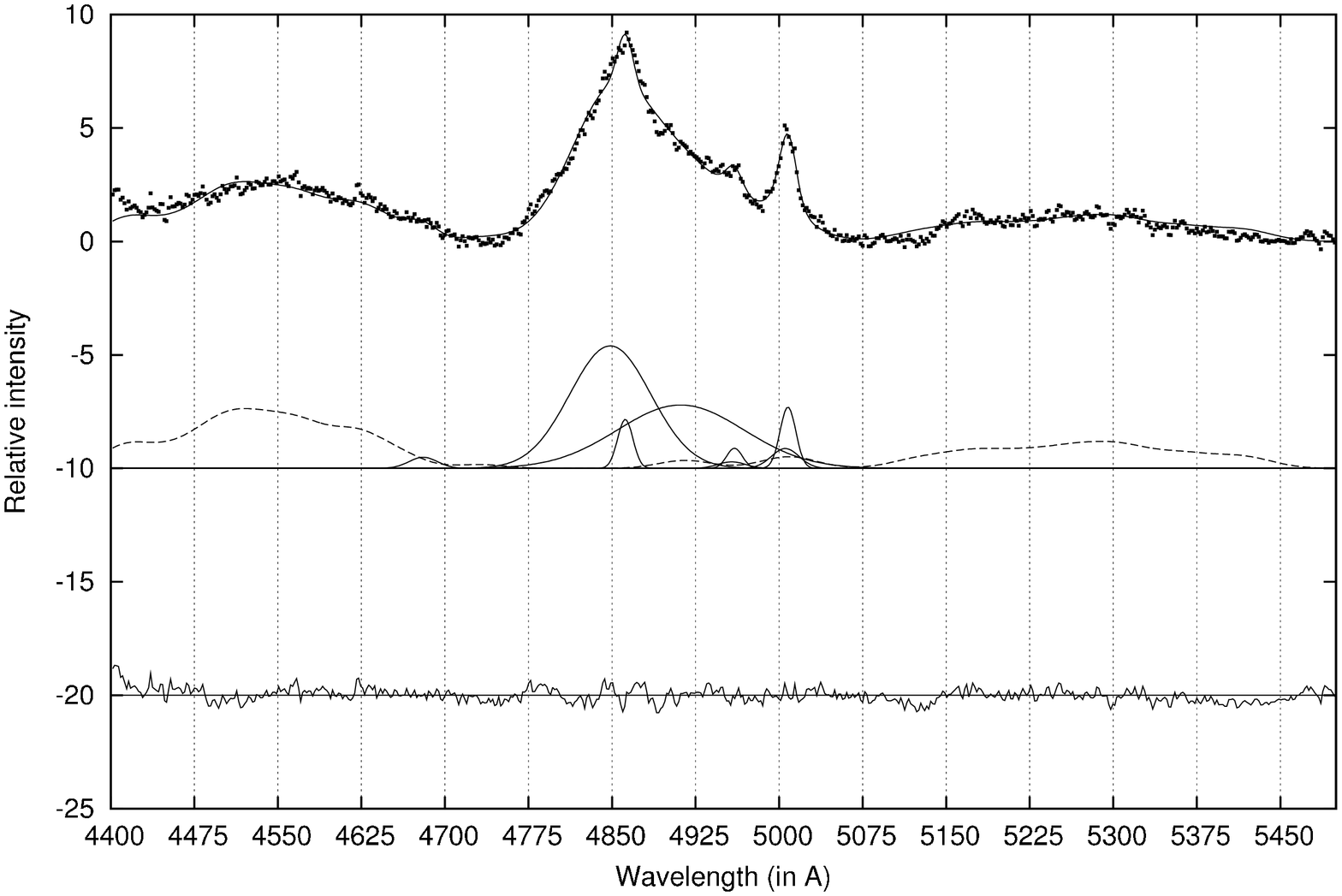}
\includegraphics[height=0.49\textwidth,angle=-90]{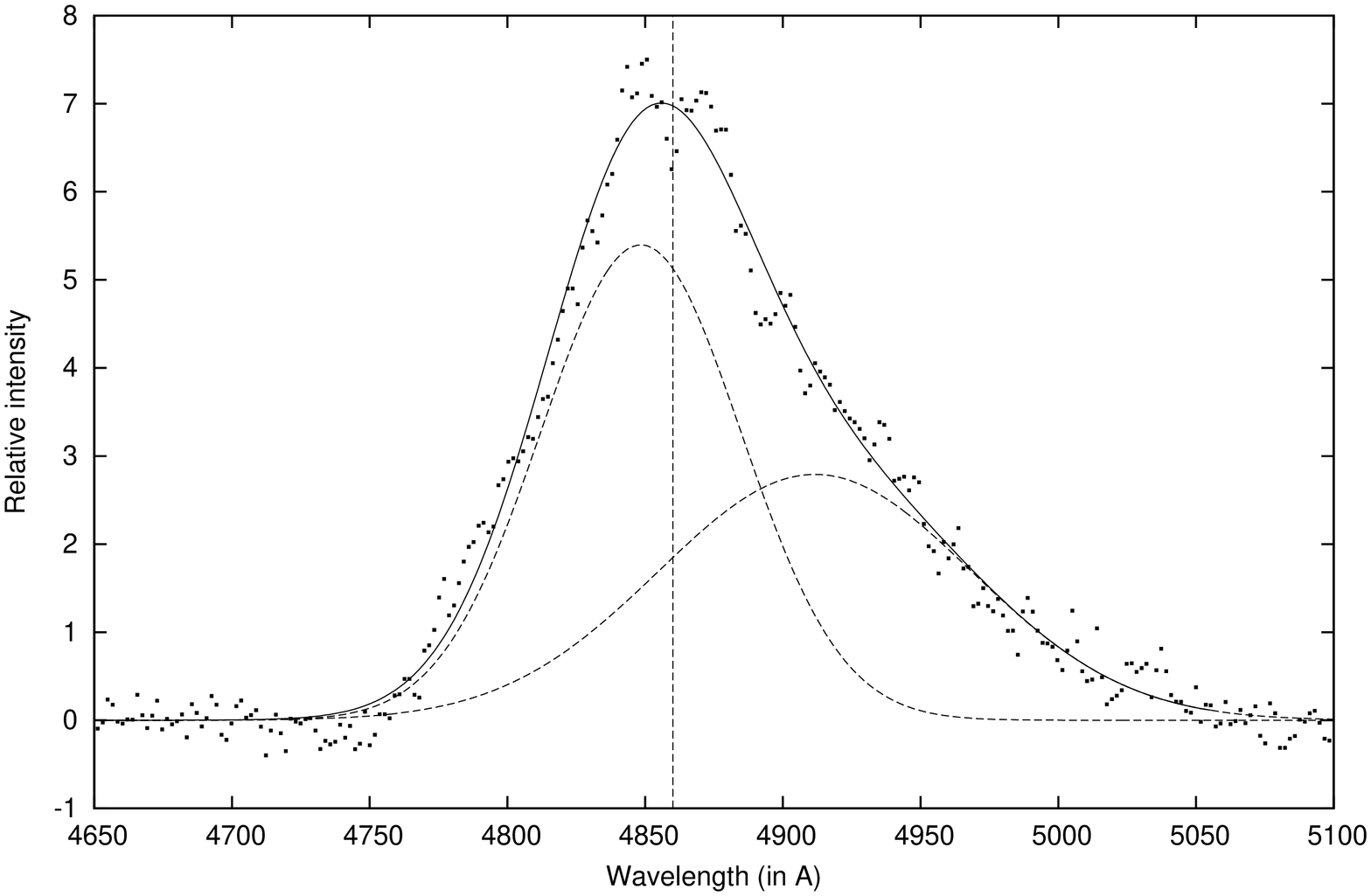}
\caption{The best fit of the H$\beta$ wavelength band (left), and
broad lines (right) after subtracting the narrow components for SDSS
J121855.80+020002.1 (up) and SDSS J162011.28+172427 (down). The vertical line
corresponds to the position of the narrow component.}
\label{fig06}
\end{figure*}

\section{Observations vs. simulations}
\label{sec:obs}

The amplitudes of the flux variations in quasars, at certain redshifts,
indicate that an enormous amount of energy is produced. The rapid flux variations
often seen are  convincing evidence of the compactness of the
emitting region. Thus, in this case a correlation between
astrometric and photometric variability will either  not exist
or be discerned only with an  astrometric precision far higher than the mas level.
At the same time, since longer, year-long, and large amplitude
variations are also recorded, the same logic would imply that the
other quasars elements are not at a standstill, as discussed. The
specific causes can be studied when and if an observed long-term,
large-amplitude optical variability is related to the astrometric
variability of the quasar photocenter \citep[][]{j03}. In addition, if this
were verified, the relationship could indicate that a large photometric
variation would make a given quasar less apt to materialize a stable
extragalactic reference frame, such as the one from the Gaia mission.
The long-term program required to monitor 
optical fluctuations in long cycles can only be established by
ground-based observations. Therefore, the astrometric limit should be on the level of few mas, which, in turn, requires high quality
seeing, telescope imaging, and relative astrometry.

We now present observations of the photocenter
variability of two objects and discuss the possibility that it was
caused by  changes to the inner structure of the AGNs.

\subsection{Observations}

{To maximize the chances  of the
photocenter variability being detected on a mas scale, $20$ quasars were selected based on
their long variability timescales and large photo-variability.
Most objects were collected from \cite{t00}, as well as \cite{macc87} and inspections of
 light curves in \cite{s93}. The observations were performed under the Observat\'orio Nacional/MCT, Brasil, telescope time
contracted to ESO at the Max Planck $2.2$ m telescope at La Silla,
Chile. The program started on April $2007$ and lasted until July
$2009$, with observations taking place about every two months.}

{The ESO2p2 WFI direct image camera is an array of 2 $\times$
4 CCDs, each covering a field of $7.5$ $\times$ $15.0$ arcmin, to
scale of
$0.238$ arcsec/px. For all the observations the same CCD was used,
keeping the quasar on a clean spot, at about one third of the diagonal
starting from the optical axis.} The same configuration was repeated
for all the observations of a same quasar, jittering allowed. The
observations as a rule were made within two hours of hour angle. Red
(Rc/162, peak $651.7$ nm, FWHM $162.2$ nm) and blue (BB\#B/123, peak
$451.1$ nm, FWHM $135.5$ nm) filters were used for each run.
Depending on
the quasar magnitude, typically from three to five images were taken
with
each filter. The integration times were never longer than $30$ min.,
yet as long as possible to provide good imaging data of the target and 
the surrounding stars. The combined signa-to-noise ratio was always close to
$1000$ for each run.

All images were treated by IRAF MSCRED for trimming, bias substraction, flat-fielding,
bad-pixel removal and split. Typically this image processing enhances the SNR
by a factor of two. The IRAF DAOFIND and PHOT tasks are employed for the
determination of centroid and (instrumental) magnitudes, with the
entry parameters adjusted for each frame. Centroids and fluxes are
obtained  adjusting  bi- dimensional Gaussians. The inner
ring where the object counting is made and the outer ring where the
sky background is counted are variable for each object and frame,
but their ratio is kept constant. The plate scale and frame
orientation are derived by IRAF IMCOORDS from positions of UCAC2 catalogue stars
(though, since the astrometry is totally relative, their values are
of no great consequence to define the correlation under study).

{Additional aspects of the method  described above were
presented in \cite{a09}, for the error
analysis, and in \cite{a11}, for relative
astrometry to derive mas-level variations. A full analysis of the
program itself will be presented elsewhere \citep{a11}.
Here we present the preliminary results regarding the R filter,
where the WFI sensitivity is higher, for two selected sources (see
Table \ref{tab03}), to exemplify the effects discussed in this
paper.} The
relative astrometric and variability procedure initially adjusts the
frames one on top of the other, in terms of coordinates and magnitudes, with respect to
the quasar position. Next, frame after frame, on the basis of the PHOT
data, the objects common to all frames are stored, provided that
the (X,Y) coordinates and the magnitudes do not vary above a chosen
threshold. {The common objects (X,Y) coordinates and
magnitudes are then adjusted by a complete third degree polynomial to
a mean frame, where   C represents either
for X, Y or M, given by}

{
\begin{eqnarray}
C_{n}^{m} - <C>_{n} =3D C_{0} + \sum_{i,j,k}^{1,3} A_{i,j,k}^{m} X^{i}Y^{j}M^{k}
\label{eqn:C}
\end{eqnarray}
}

Finally, a further round of analysis discards the reference objects
for which (X,Y) or magnitude variation are above the threshold. The
averages of (X,Y) and magnitude for the remaining objects (with reference
to the quasar as a fixed origin) are obtained and correlated with both
the time-line and  each other.

{Table \ref{tab03} presents the timeline variation in position
and
magnitude for quasars J121855.80+020002.1 and J162011.28+172427.5,
with reference values brought from the CDS. The quantities of final
comparison stars were 8 for quasar
 J121855.80+020002.1, and 30 for quasar J162011.28+172427.5, where we
note that for the initial frame-to-frame adjustment the number of
stars used was always much larger. As a consequence, the positional
errors had a mode of $1.5$ mas for the first object and $15$ mas for
the second - whereas the magnitude errors had mode $0.001$ for both
objects.}

\begin{table}
\caption{. The summary of the measurements of the offset of
photocenter: Col. 1 - the mean epoch of observation; Col. 2 - the
time interval in days between each measurement; Col. 3 - the X-direction
(basically RA) astrometric variation in mas from the previous
measurement ; Col. 4 - the Y-direction (basically DEC) astrometric
variation in mas from the previous measurement; Col. 5 - the
magnitude variation given in tenths of magnitude from the previous
measurement. In the first lines, the values correspond to the offsets
to the nominal CDS references. In the subsequent lines, we present the
offsets to the previous line values. The combined corresponding errors ($\sigma$) are given.}
\begin{tabular}{c r r r r}
\hline
\hline
~ \\
\multicolumn{5}{c}{
SDSS J121855.80+020002.1, z = 0.327, MAG$_R$ = 18$^m$.1} \\
~ \\
\hline
\hline \\
DATE & DAYs & $\Delta$RA$\pm\sigma$ & $\Delta$DE$\pm\sigma$ &
$\Delta$MAG$_R\pm\sigma$ \\
 &  &  (mas) &  (mas) &  (10$^{-1}$)\\
\\
2008.016 & 0.0   & -11$\pm$3 & -~3$\pm$2 & -0.420$\pm$0.009 \\
2008.163 & 53.4  & +15$\pm$3 & +~4$\pm$2 & +0.134$\pm$0.007 \\
2008.263 & 36.8  & +~6$\pm$2 & -~3$\pm$1 & -0.917$\pm$0.011 \\
2008.415 & 55.4  & -~4$\pm$1 & +~1$\pm$1 & +1.774$\pm$0.012 \\
2008.970 & 202.7 & -~8$\pm$1 & +~4$\pm$1 & -1.773$\pm$0.007 \\
2009.382 & 150.5 & ~~0$\pm$4 & -~3$\pm$2 & +2.070$\pm$0.011 \\
\hline
\hline \\
\multicolumn{5}{c}{SDSS J162011.28+172427.5, z = 0.112, MAG$_R$ = 16$^m$.2} \\
~\\
\hline
\hline \\
DATE & DAYs & $\Delta$RA$\pm\sigma$  & $\Delta$DE$\pm\sigma$  & $\Delta$MAG$_R\pm\sigma$ \\
 &  &  (mas) &  (mas) &  (10$^{-1}$)\\
\\
2007.277 & 0.0   & -17$\pm$10 & +24$\pm$22 & -0.136$\pm$0.007 \\
2007.430 & 58.8  & +~2$\pm$9 & ~~0$\pm$24 & +0.032$\pm$0.007 \\
2008.415 & 356.9  & +14$\pm$6 & +~4$\pm$21 & +0.021$\pm$0.006 \\
2008.647 & 84.9  & -23$\pm$8 & -59$\pm$16 & +0.002$\pm$0.005 \\
2009.181 & 195.1 & +76$\pm$19 & +36$\pm$15 & +0.550$\pm$0.023 \\
2009.382 & 73.3 & -47 $\pm$18 & -58$\pm$22 & -0.560$\pm$0.023 \\
\hline
\hline
\end{tabular}
\label{tab03}
\end{table}

{From the six values of right ascension and declination
variation for the two example sources in Table \ref{tab03}, we can
calculate
the non-parametric correlations against the magnitude variations.
They were calculated by the Spearman rank correlation coefficient,
which permitted weighting by the inverse squared sum of the
 position and magnitude uncertainties. For quasar
J121855.80+020002.1, the correlations are
$\Delta$RA$\times$$\Delta$MAG=0.44 (significance 0.03)\ and \ 
$\Delta$DE$\times$$\Delta$MAG=0.56 (significance 0.01). For quasar
J162011.28+172427.5, the correlations are
$\Delta$RA$\times$$\Delta$MAG=0.75 (significance 0.01) \ and \
 $\Delta$DE$\times$$\Delta$MAG=0.75 (significance 0.01). Consequently, there is no  significant correlation between the photocenter  and magnitude
 variation (significance $>> 10^{-5}$).}

\begin{figure}
\centering
\includegraphics[width=8cm]{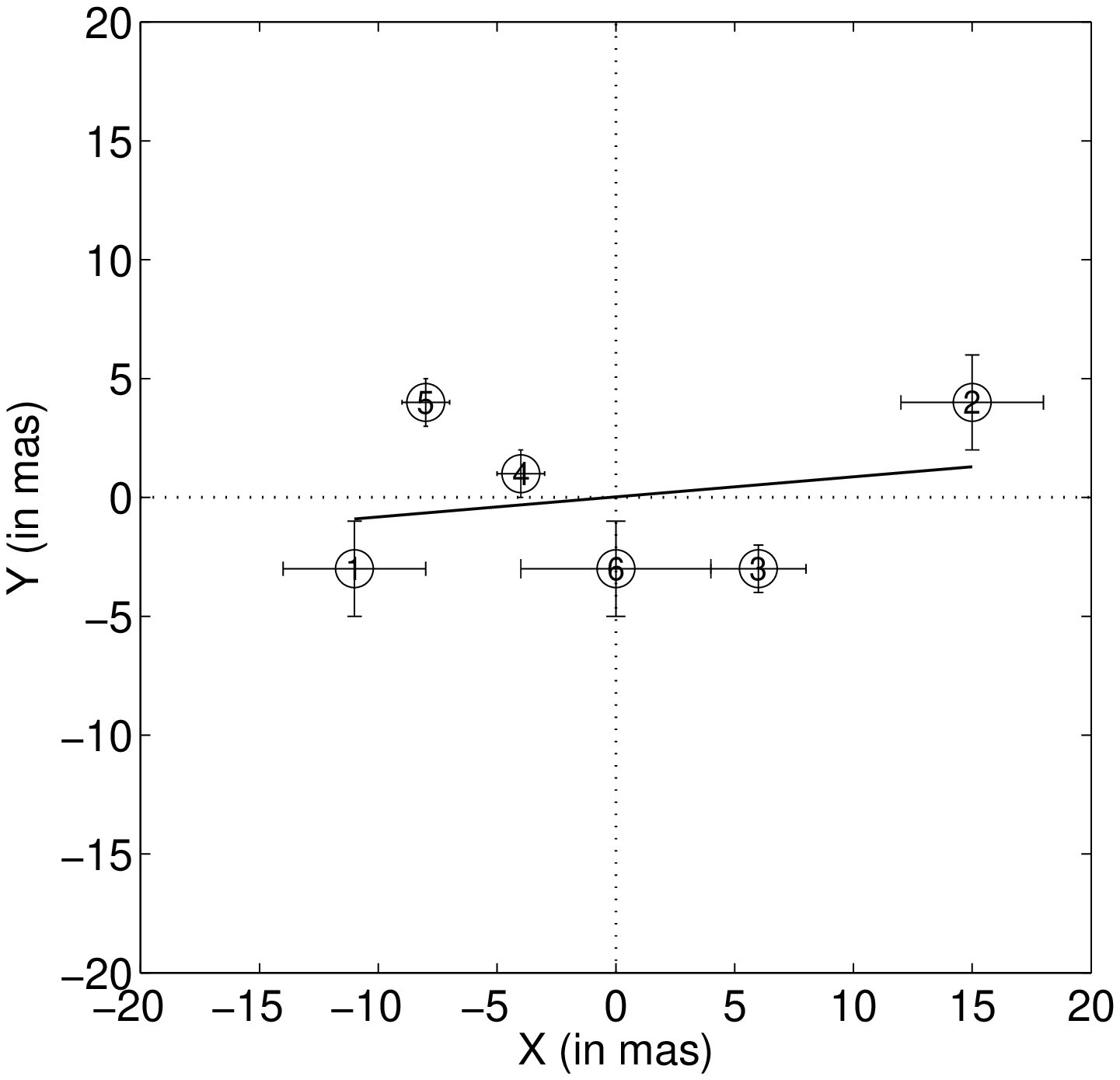} \\
\includegraphics[width=8cm]{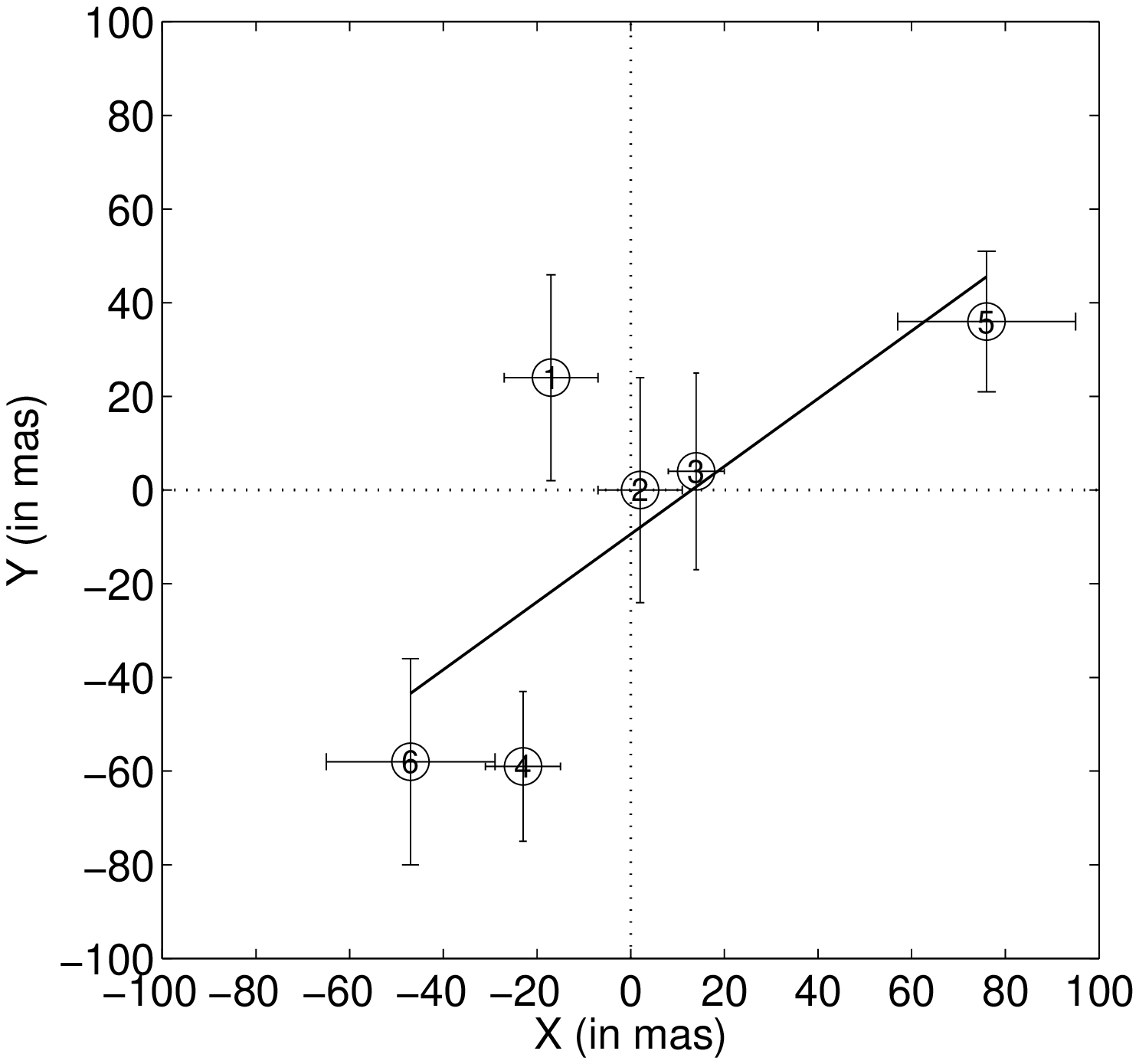}
\caption{ Observed  astrometric
variability of the photocenter, measured for SDSS
J121855.80+020002.1 at $z=0.327$ (up) and SDSS J162011.28+172427.5 
(MRK 877) at $z=0.112438$ (down), during the period 2007-2009
(see Table \ref{tab03}). 
The straight solid lines in the  panels represent linear fits
through the observed positions. Notations of points from 1 to 5
correspond to the observational epochs from first to the last as
given in Table \ref{tab03}.} \label{fig07}
\end{figure}

\subsection{Comparison between the simulated and the observed variations}

To explore whether the observed variations in SDSS
J121855.80+020002.1 ($z=0.327$, $18^m.1$) and SDSS
J162011.28+172427.5 (Mrk 877, $z=0.112438$, $16^m.2$) are caused by
perturbations in the accretion disk\footnote{Both observed
objects have broad lines (type 1 AGN); in our simulations, we found that the
photocenter offset is significant only when the central source is
partly obscured by the dust. Therefore, there is a small chance
that the observed variations are caused by changes in the torus
structure.},  we first estimate the masses of the black holes ($M_{bh}$)
for these two objects. There are several estimators for $M_{bh}$ in
AGN \citep[see e.g][and reference therein]{m08}, and to measure them
for these two objects we used spectra observed with HST
(for SDSS J121855.80+020002.1) and from SDSS database (for SDSS
J162011.28+172427.5). We first measured from spectra the luminosity
at 5100 \AA\ and decomposed spectra using a multi-Gaussian fit
\citep[see e.g.][]{p04}. In Fig. \ref{fig06}, the best fit and the
broad component after subtraction of the narrow and Fe II lines are
shown. As  can be seen in Fig. \ref{fig06}, the broad H$\beta$ line
in both objects has a red asymmetry, indicating a very complex
geometry of the BLR. Also, two separated broad components may
indicate the presence of disk emission. After measuring the full
width at half maximum (FWHM), we used the  three estimators
$M_{S}$, $M_{V}$ and $M_{N}$, given by \cite{s03}, \cite{v06}, and
\cite{n07}, respectively. The estimated masses for SDSS
J121855.80+020002.1 are: $M_{S}=9.18\times 10^8\ M_\odot$,
$M_{V}=1.37\times 10^9\ M_\odot$, and $M_{N}=1.10\times 10^9\ M_\odot$,
or on average $M_{bh}=(1.13\pm 0.23)\times 10^9\ M_\odot$. In the
same way, we estimated the black hole masses of SDSS
J162011.28+172427.5 to be $M_{S}=3.51\times 10^8\ M_\odot$,
$M_{V}=5.25\times 10^8\ M_\odot$, $M_{N}=3.72\times 10^8\
M_\odot$, or on average $M_{bh}=(4.16\pm 0.95)\times 10^8\ M_\odot$.

{To estimate the possibility that the photocenter
variability is caused by some perturbation in the disk (or in the
BLR), we calculated dimensions of the BLR of these two objects, using
the relation between the BLR radius and luminosity at $5100$ \AA\
\citep[see e.g.][]{v06}. We estimated the BLR sizes for SDSS
J121855.80+020002.1 to be around $113$ light days (that is   $\sim 0.02$ mas)
and for Mrk 877 $10$ light days ($\sim 0.004$ mas). Therefore, the observed
photocenter variability cannot be explained by the perturbation in
the BLR.}.

\subsection{Possible explanation of the photocenter variability in
SDSS J121855.80+020002.1 and Mrk 877}

As we noted in \S 5,2, a perturbation in the accretion
disk cannot explain the photocenter jitter observed in the two
quasars. Moreover, we have estimated that the BLR in both objects is very
compact, around 10$^{-5} - 10^{-6}$ arcsec  (that
translates into  light day to several hundred light day scale), which
is inconsistent with  the photocenter variations. We note that
these compact regions cannot be resolved by Gaia, as its PSF will be 
$\sim 200$ mas.

For  objects that are partially obscured, a variation in both the
central luminosity and  the dust sublimation radius may produce an
offset in the photocenter (at z=0.1, see Table \ref{tab:tor}), of
about of one tenth arcsec. However it cannot explain the photocenter of
the two quasars under study, as the jitter is  smaller, and 
 they both exhibit broad emission lines, which implies that they have a
geometry where obscuration is very small or nonexistent. 

Another possible source of  photocenter variability are ``nuclear"
super-novae.
Several studies \citep[see e.g.][etc.]{c04,d07,p09} demonstrate that AGNs
may be associated with star formation regions. For instance \cite{d07} found
that on kpc (or pc)  scales (corresponding to the observed
photocenter variation in our objects) the luminosity of
the starburst component may be comparable to that of the AGN.\\

{
For the (U)LIRGs (ultra luminous infrared galaxies), the
expected
supernova rate is very high, as high as  $2.4$ yr$^{-1}$, if the
infrared luminosity is  produced entirely by starbursts
\citep[see]{man03}. In
the extreme case of this kind of objects, a large supernova rate
(SNr) may have
influence on the stability of the photocenter.  We estimate the SNr,
considering the relation given in \cite{mm01}, and assuming that
 the SNr and   the star formation  rate (SFR) are correlated
\citep{man03},
 the latter calculated using the luminosity of the
H$\alpha$ line \citep{cal07}.
We could only calculate the SNr  for  SDSS J121855.80+020002,
because we do not have H$\alpha$ spectral data for Mrk 877. We obtained
SFR$\approx$14.7 yr$^{-1}$ and a corresponding SNr$\approx$0.1 yr$^{-1}$ (i.e. one SN
every ten years)
for SDSS J121855.80+020002. We conclude  that it is unlikely that
supernovae
are  responsible for the photocenter shift of this object. \\}

{We now discuss a scenario where the photocenter jitter might be
related to the jet emission.  In terms of radio loudness
\citep[][]{kell89}, i.e. $R=F_{5GHz}/F_{B band}> 10$,
SDSS J121855.80+020002.1 has a value of $R \sim 1.1$ \citep[][]{raf09}, and Mrk 877
has 
$R \sim 0.41$ \citep[][]{sik07}, very far from the values shown by
radio loud
quasars, which tend to have relativistic jets. Radio-quiet objects can have jet emission \citep[e.g. Mrk 348,
see][]{ant02},
though their radio-brightness can be significantly higher
\citep[][]{ant02} than that of the objects under study. There are VLA
$1.4$GHz  maps
at the position of our sources. The  FIRST map of SDSS
J121855.80+020002
shows a faint core-morphology on the $1$ mJy level, and  in the case of
Mrk 877 there is no detection with NVSS at the position
of the optical source. We conclude that there is no evidence
 that the jet plays a role in the photocentric  variation of these
objects.
}

It is interesting that in the two objects (see Fig.
\ref{fig07}) the photocenter offset is {almost}
aligned,
especially in  SDSS J121855.80+020002,  with a straight
line.

These aligned positions of the photocenter
offset may correspond to  two variable sources close to each other,
with the photocenter always shifting towards the brighter of the
two.
A speculative possibility is a binary supermassive black-hole
system,
of the type discussed in \citep[see
e.g.][and references therein]{lb09,bO09,sh10,b11,pg11},
and  based on the observations of double-peaked narrow and broad
lines.
 We note that the broad-line shapes
of the objects under study are complex (see Fig. \ref{fig06})
and can be properly fitted with two broad
Gaussians that are shifted (toward either the blue or red) with respect to the
central narrow component (the vertical line in Fig. \ref{fig06}.
In \cite{pop00} and  \cite{sl10},  a
binary broad emission-line region has been investigated, and the line
profiles of  this system have been discussed. To detect two peaks in
the broad line profile, it is necessary 
to be able to resolve the two BLRs, and  the
 plane of the orbit must be edge on with respect to the line
of observation. An asymmetric line profile might
 result solely  from a system  where the two
BLRs have different dimensions and luminosities
 \citep[see Figs. 4-8 in][]{pop00}.  Such a system might exist at the
center of
our quasars, and  may be the cause of their photocenter
variability.

{We note that in addition to  the binary black hole scenario, the
superposition of
two visually  close and variable sources  \citep[see the several examples presented
in][]{pg11} can explain an aligned variability. All of these scenarios
should be considered in future investigations.}

\section{Conclusions}
\label{sec:conc}

We have simulated the perturbation in  the inner structure
of quasars (accretion disk and dusty torus),  to find how
much these effects can offset their photocenters, and try to determine whether
it will be observable with future Gaia mission. We have
considered two AGNs whose the photocenter variations have been  observed,
in order to compare them with our simulations. From our
investigations, we draw the following conclusions:

i) Perturbations (or bright spots) in an accretion disk may cause an
offset of the photocenter, and this effect has a good chance
 of being detected by the  Gaia
mission. The most likely candidates are low-redshifted AGNs with massive
black holes (10$^9$-10$^{10}$) that are in principle very bright
objects. One can expect a maximal offset of the center (in the case
of a bright spot located at disk-edge) on the order of few mas.

ii) A photocenter offset can be caused by changes to the torus structure
due to different illuminations of the torus  when the
central source is obscured by the dust. A maximal offset can be
several mas, which also  be  detectable with Gaia.

iii) A photocenter offset caused by both effects is connected to the
photometric variation in the objects, but there is a small
probability of a correlation between astrometric and photometric
variations. We note here that quasars with high photometric variability
are not good objects for constructing  the optical reference
frame.

iv) To exclude the possibility of the photocenter variation being caused by a
perturbation in the accretion disk, or in the BLR, one may estimate the
dimensions of the BLR and choose objects with a compact BLR. However,
 to avoid any variation in the photocenter caused by filaments
in the torus, it is preferable to choose quasars with face-on
oriented tori.

v) The observed photocenter variability of two quasars cannot be
explained by the variation in their inner structure (accretion disk
and torus). It seems that the observed photocenter variation can be reproduced very
well by a scenario with double variable sources at the center of these
objects. It may indicate (as well as complex broad line shapes) that
these objects are good candidates for binary black hole systems.

At the end, we conclude that Gaia, in addition to providing astrometrical
measurements, may be very useful for an astronomical investigation of
the inner quasar structure (physical processes), especially in low
redshift variable sources.

\begin{acknowledgements}
This research is part of the projects ''Astrophysical Spectroscopy
of Extragalactic Objects`` (176001) and ''Gravitation and the Large
Scale Structure of the Universe`` (176003) supported by Ministry of
Education and Science of the Republic of Serbia. L.\v C. P., A. S.
and P. J. are grateful to the COST action 0905 "Black Hole in an
Violent Universe" that helps them to meet each other and discuss the
problem. M. S. acknowledges support of the European Commission
(Erasmus Mundus Action 2 partnership between the European Union and
the Western Balkans, http://www.basileus.ugent.be) during his
mobility period at Ghent University. S. Ant\'on acknowledges support
from
FCT through Ciencia 2007 and PEst-OE/CTE/UI0190/2011. The European
Commission and
EACEA are not responsible for any use made of the information in this
publication. {We would like to thank an anonymous  referee for very helpful comments.}
\end{acknowledgements}

\end{document}